\documentclass[runningheads]{llncs}
\usepackage{graphicx}
\usepackage{float}
\usepackage{geometry}
\usepackage{comment}
\usepackage{csquotes}
\usepackage{makecell}
\usepackage{lineno}
\usepackage{amsmath,amssymb,amsfonts} 
\usepackage{algorithmic}
\usepackage[linesnumbered,ruled]{algorithm2e}
\usepackage{float} 
\usepackage{textcomp} 
\usepackage{eso-pic}

\usepackage{xcolor}
\usepackage{subfigure}
\usepackage[english]{babel}
\usepackage[utf8]{inputenc}

\usepackage[textsize=footnotesize, textwidth=1.2cm]{todonotes}

\begin{document}
%

\title{Indoor and Outdoor Crowd Density Level Estimation with Video Analysis through Machine Learning Models}

\titlerunning{Machine Learning-based Indoor and Outdoor Crowd Density Level Estimation with Video Analysis}

\author{Mahira Arefin\inst{1} \and Md. Anwar Hussen Wadud\inst{1, 3} \and
Anichur Rahman\inst{2, 3} \orcidID{0000-0002-2691-1748}\thanks{Corresponding author}}

\authorrunning{Mahira Arefin et al.}
%
\institute{Bangladesh University of Business and Technology, Dhaka, Bangladesh\\ 
\email{mahiraarefin19@gmail.com},
\email{mahwadud@gmail.com}
\and National Institute of Textile Engineering and Research, constituent institute of University of Dhaka, Nayarhat, Savar, Dhaka-1350, Bangladesh\\
\email{anis\_cse@niter.edu.bd}
\and Mawlana Bhashani Science and Technology University, Tangail, Bangladesh
}

\maketitle              

\begin{abstract}


Crowd density level estimation is an essential aspect of crowd safety since it helps to identify areas of probable overcrowding and required conditions. Nowadays, AI systems can help in various sectors. Here for safety purposes or many for public service crowd detection, tracking or estimating crowd level is essential. So we decided to build an AI project to fulfil the purpose. This project can detect crowds from images, videos, or webcams. From these images, videos, or webcams, this system can detect, track and identify humans. This system also can estimate the crowd level. Though this project is simple, it is very effective, user-friendly, and less costly. Also, we trained our system with a dataset. So our system also can predict the crowd. Though the AI system is not a hundred percent accurate, this project is more than 97 percent accurate. We also represent the dataset in a graphical way. 

\vspace{2mm}
\end{abstract}

\section{Introduction}
Nowadays, there is a need to detect the crowd, track them and estimate the crowd level in various places. This is important for safety purposes and public services. There are many datasets that exist to train the system to predict the crowd, and there are many models to train the system. Various models help to increase the accuracy of the system. Using graphical representation helps to understand the system and datasets \cite{khan2022accurate}, \cite{Debnath2022}. There are many crowd detection and estimation systems. An author has approached the Real-time Crowd Motion Analysis system \cite{n1}. Here examining motion patterns was used to complete this project. It can detect sudden changes of motion in a high-density crowd. In the future, more iterations can increase the accuracy of motion detection.
Another author has approached the speedy and Accurate Detection and Localization of Abnormal Behavior in Crowded areas \cite{n2}.  Here the proposed methodology provided a cubic-patch-based approach using a source of classifiers, which assemble the result of two different types of video descriptors \cite{khan2022multinet}. This project shows comparable accuracy in prediction and time efficiency to state-of-the-art methods, but this AI project is complex and requires more research \cite{rahman2022federated}, \cite{9499121}. 

\vspace{2mm}
Moreover, another author has approached the Geographical Regularities of Crowd Behaviors for a social media-based Geo-social outcome detection\cite{n3}, \cite{rahman2022sdn}. Here they used geographical regularities derived from crowd behaviour patterns to identify abnormal geo-social incidents. This project can identify local events, such as festivals or concerts, by analyzing Twitter posts that are written about these events, but this project can perform just on Twitter. Hence, we found some limitations in the existing systems. Firstly, most of the systems are not cost-effective. Secondly, many of the systems are only suitable for less dense areas. Thirdly, the Accuracy level of the prediction is low. Fourthly, some projects are based on particular social media. Lastly, most of the system does not have an estimated crowd level\cite{RahmanWiley} \cite{rahman2024machine}. So, we built an AI project that is very user-friendly and cost-effective. It can be used in any environment, and the accuracy rate is also high. This project also proves the crowd estimation level. It counts the total number of people and the current number of people, making this project unique from similar projects. This paper’s contribution is--
 
\begin{itemize}
 \item We proposed an AI project using a Caffe model and Linear regression algorithm. \vspace{1mm}

\item This paper also counts Frames Per Second (FPS).\vspace{1mm} 

\item Graphical representation of the dataset makes this project easily understandable.  \vspace{1mm}

\item The tracking frame waits a few seconds for the person who has left the screen. 
\end{itemize}


\textbf{Organization:} The rest of the paper is organized: Segment 2 describes the related works with recent studies. In segment 3, the authors describe the proposed methodology, and it describes them. Section 4 describes the details of the Experiments and discussion, and finally, section 5 concludes the study.

\section{Related Works}
Several researchers proposed in recent years based on crowd density level estimation through
machine learning models. In this part, we are going to
discuss some literature review of recent works:
\newline
\newline
Almeida et al. approached Change detection in populated areas. Here they provide a method for spotting unusual behaviour in crowds of people using histograms of velocity in world coordinates. A mixture to remove the background with the flow is used to extract the overall motion at each image frame, removing unimportant motion vectors brought on by noise, non-stationary background pixels, and compression issues. In the future, they propose the use of Ht(i, j) to create 1D histograms exclusively for speed and orientation \cite{n4}. Gray et al. approached A Crowd Counting Approach with Viewpoint Invariance. Here they provide a learning-based method that uses a single camera to count people in crowded spaces independent of the camera's perspective or orientation. The experimental outcomes, which encompass various sites with different camera orientations, show the efficacy and potential of the suggested technique. In this implementation, the blob size histogram is separated into six bins with a size of 500 and consistent spacing, and the edge orientation histogram is divided into eight bins spanning from 0 to 180 degrees \cite{n5}. Cao et al. approached the suggested method in this study using crowd motion features, such as crowd kinetic energy and motion directions, to detect anomalies in crowded settings. The approach determines abnormal behaviour based on departures from the anticipated crowd behaviour by calculating the crowd's kinetic energy and motion directions. They plan to use high-performance hardware to enhance the system's performance in the future, such as graphics processing units (GPUs)\cite{n6}.
Foggia et al.  presented using acoustic analysis to find car crashes on busy highways. The method uses audio signal analysis over short and long periods to find both brief and extended occurrences. To increase the accuracy of event detection, the authors suggested a method that combines short- and long-term analysis of acoustic signals. Additionally, the scientists evaluated a road traffic noise model to determine how well the system performed for sound sources at various distances and presented an architectural deployment method\cite{n7}.

\vspace{2mm}
On the other study, Shen et al. focused on Behaviour-aware group detection in crowded urban environments with WiFi probes known as BaG. This study suggests a group identification system dubbed BaG that uses smartphone usage patterns and mobility data to identify and analyze group behaviour. They suggest employing collective matrix factorization to expose hidden relationships in their proposed detection approach. This is accomplished by concurrently taking into account usage patterns and mobility information. But this project is complex, and it needs highly capable resources\cite{n8}. Subburaman  et al. covered the Crowd Size Calculation Using a Common Head Detector. The head is often the most visible feature of the body in such circumstances; hence the authors of this research suggest a method for counting people by detecting their heads in crowded scenes. They employ a cutting-edge cascade of enhanced integral features as their head detector to do this. By lowering false detections and introducing techniques to account for occlusions in future work, the scientists also hope to enhance the system\cite{n9}. Arandjelovic  et al. presented the Detecting crowds in still images. The study discusses the issue of crowd detection, which has numerous implications for traffic control, urban planning, and law enforcement. This appearance-based approach examines the distribution of quantized SIFT words over a picture using a statistical Poisson model; they still have to work on its accuracy level\cite{n10}. Chen et al. focused on the  Crowd Behavior Analysis Based on the Acceleration Feature. In this study, a novel method for identifying unusual crowd behaviour in video surveillance systems is presented. The technique uses an acceleration-based algorithm to classify behaviour and is based on the grey-scale invariance of three adjacent frames. In the future, they proposed work with high-density crowd scenes\cite{n11}. Gavrilis et al. addressed using decoy hyperlinks to detect flash crowds. The use of dummy hyperlinks in this paper's detection of Denial of Service (DoS) attacks on web services is innovative. In comparison to previous methods, such as graphic Turing tests, the suggested method for detecting Denial of Service assaults utilizing decoy hyperlinks has a number of advantages. It is simple to use and doesn't need any particular user interaction. Decoys have also been embedded in a small subset of website pages using an algorithm that has been proposed\cite{n12}, \cite{9528133}.
\vspace{2mm}

Arulanandam et al. discussed the Detection of Abnormal Events in Crowded Video Scenes. To extract features from video frames, the proposed system combines a number of image processing methods, such as the DoG filter, MHOF, and EOH. The method seeks to provide a novel way for anomaly identification and is created to function for busy video sequences. In the future, they seek to create a system that can process the entire video for anomaly detection and enhance system effectiveness when handling greater patch sizes by adding additional techniques\cite{n13}, \cite{rahman2021smartblock}. Shahzad et al. covered the Smart Surveillance System with Multi-Person Tracking for Crowd Counting and Normal/Abnormal Events Detection. The system is made up of four modules: crowd cluster analysis, tracking, head-torso template extraction, and people detection. People are recognized using head-torso cues, and human silhouettes are recovered using an inverse transform and median filter. Each individual is tracked using Kalman filter techniques with Jaccard similarity and normalized cross-correlation. In the future, they plan to improve accuracy in event detection\cite{n14}. Thida et al. presented the Local Abnormality Detection in Crowded Scenes Using a Laplacian Eigenmap with Temporal Constraints. By examining local motions' spatial and temporal fluctuations in an embedded environment, they present a spatiotemporal Laplacian eigenmaps approach that extracts distinct crowd behaviours. The training and testing phases of a new framework for locating anomalous regions in crowded scenes are presented in this research, but this project takes more time to estimate\cite{n15}. Mousavi et al. discussed Detecting Abnormal Crowd Behavior by Analyzing Tracklets. In order to identify unusual circumstances in crowded images, the study offers a brand-new video descriptor called Histogram of Oriented Tracklets. The proposed strategy is based on tracks, which are long-range motion trajectories, as opposed to conventional methods that use optical flow, but the project is quite complex\cite{n16}, \cite{hasan2021normalized}.

\section{Proposed Methodology for Crowd Density Level Estimation with Video Analysis}

\subsection{Data Collection and Device Management}
\begin{figure*}[h]
\centering
\centerline{\includegraphics[height=7cm]{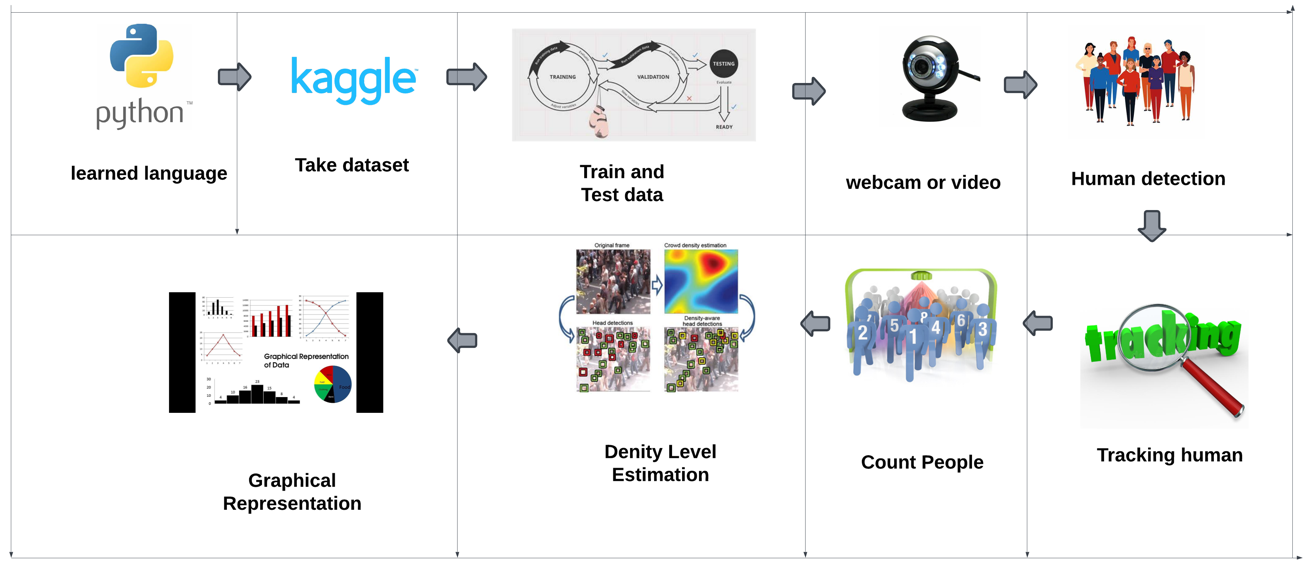}}
\caption{System Model for Crowd Density Level Estimation}
\label{fig:r1}
\end{figure*}

In Fig. \ref{fig:r1}, we cover the methodology of our system here. At first, we learned Python programming language. Then we took an organized crowd-density dataset. After that, we trained and tested our system through the dataset. Our project can detect humans from webcams, images, or any videos. Also, it can track humans and count people. Then our project estimates the density level of the crowd. Lastly, using graphical representation, we can easily understand the dataset and the whole manuscript\cite{rahman2024internet}. 

\subsection{Flowchart with Technical description}
\textbf{Flowchart:}
\begin{figure*}[!ht]
\centering
\centerline{\includegraphics[height=15cm]{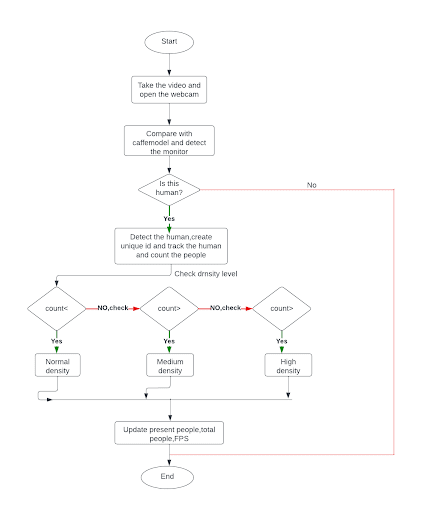}}
\caption{Flowchart of the proposed system}
\label{fig:r2}
\end{figure*}
Here in Fig.\ref{fig:r2},  at first our system takes videos and a webcam compares the object with the Caffe model to recognize if it is human or not. if no humans exist, it will terminate. if it detects any human here, it tracks people with unique IDs and counts the people. It also estimates the crowd density level according to the number of people. Lastly, system updates present people, total people, and fps.

\vspace{2mm}
\textbf{CODE Segments: }
\newline
\textbf{For image detection:}
\begin{verbatim}

protopath = "MobileNetSSD_deploy.prototxt"
modelpath = "MobileNetSSD_deploy.caffemodel"
detector = cv2.dnn.readNetFromCaffe(prototxt=protopath, caffeModel=modelpath)

CLASSES = ["background", "aeroplane", "bicycle", "bird", "boat",
           "bottle", "bus", "car", "cat", "chair", "cow", "diningtable",
           "dog", "horse", "motorbike", "person", "pottedplant", "sheep",
           "sofa", "train", "tvmonitor"]

def main():
  

    
        if confidence > 0.5:
            idx = int(person_detections[0, 0, i, 1])

            if CLASSES[idx] != "person":
                continue    

\end{verbatim}
Here, we declared the photo path and model path at first. then, declared the classes of various objects and persons for identification. Then compare the person with the model. If it matches more than 50 percent then it will assume that this is a human.

\textbf{For video detection:}
\begin{verbatim}
def main():
    cap = cv2.VideoCapture(0)
   
            if confidence > 0.5:
                idx = int(person_detections[0, 0, i, 1])

                if CLASSES[idx] != "person":
                    continue


            fps = (total_frames / time_diff.seconds)


\end{verbatim}

Here, we declared the photo path and model path at first. then declared the classes of various objects and persons for identification. Then it takes the webcam video. Then it compares the video with the person from the model. Then it calculates the fps.

\vspace{2mm}
\textbf{For people counter:}
\begin{verbatim}


def main():
    cap = cv2.VideoCapture('a.mp4')

            if confidence > 0.5:
                idx = int(person_detections[0, 0, i, 1])

                if CLASSES[idx] != "person":
                    continue

            fps = (total_frames / time_diff.seconds)

       
        if lpc_count < 15:
            text = "normal crowd"
            cv2.putText(frame, text, (80, 300),
            cv2.FONT_HERSHEY_COMPLEX_SMALL, 2, (0, 150, 255), 3)

        elif 15<= lpc_count <= 25:
            text = "medium crowd"
            cv2.putText(frame, text, (80, 300),
            cv2.FONT_HERSHEY_COMPLEX_SMALL, 2, (200, 0, 255), 3)

        else:
            text = "high crowd"
            cv2.putText(frame, text, (80, 300),
            cv2.FONT_HERSHEY_COMPLEX_SMALL, 2, (0, 0, 255), 3)


\end{verbatim}

Here, we declared the photo path and model path at first. then, declared the classes of various objects and persons for identification. then it counts the frames with a loop. Then it takes a pre-recorded video and compares it with the model. Also, it counts the frames. also selects the font size of the frame writing. then for the estimation it counts the people, if the number of people is less than 15 then it will show as a Normal crowd. If the number of people is equal or larger than 15 and less than 25 it shows as a Medium crowd. else it shows a high crowd.

\subsection{ Video and Crowd Analysis with Algorithm Representation: }

We used a ‘caffe model’ to recognize people from the crowd.  Here we also import the model in our program . It is a deep learning framework which is written in C++ and has Python and Matlab bindings. 
After completing the project we used “Sequential algorithm” to train the dataset and various graphical representations.
We also used Linear regression to calculate the accuracy rate and predict the data. Step by step process of the sequential algorithm is given below:

\begin{itemize}
    \item \textbf{Step-1:} When attempting to determine the relationship between two variables, the term regression is utilized.\vspace{2mm}
    
    \item \textbf{Step-2:} That link is employed in statistical modeling and machine learning to forecast how future occurrences will turn out \cite{kundu2024federated}.\vspace{2mm}
    
    \item \textbf{Step-3:} In linear regression, a straight line is drawn across all of the data points using the connection between them.\vspace{2mm}
    
    \item \textbf{Step-4:} Future values can be predicted using this line.
\end{itemize}

\section{Experiment Analysis and Discussion}
\subsection{Data Description}
An assortment of structured and ordered data used for analysis, research, and other reasons is referred to as a data set. A data set can be any size and can include a range of data types, including text, numerical, and category data.  As they offer a useful source of information for training algorithms and creating prediction models, data sets are frequently utilized in machine learning and data science applications. Additionally, they can be utilized in statistical analysis to draw conclusions or test theories about a population \cite{rahman2024blocksd}.
The dataset provides the necessary input for teaching algorithms and creating predictive models, data sets are crucial to AI. An AI system cannot be educated efficiently without data, and it will not be able to carry out its intended task. Additionally, data sets are essential for validating and testing AI models. Researchers can assess a model's accuracy and efficacy by comparing the projected output to the actual result. Additionally, data sets aid in ensuring the impartiality and fairness of AI systems. In conclusion, data sets are essential to AI since they enable the development of prediction models, the testing and validation of AI systems, and the assurance of objectivity and fairness. From the Kaggle website, we used a \enquote{crowd-counting} dataset in this instance. It is used by us for evaluation and numerous graphical displays. 

\subsection{Results Analysis}
After completing the project we can see the whole project and can evaluate the project with accuracy, graphical representation, and the various models and algorithms. In this chapter, we have shown all this portion briefly.
\newline

\begin{figure*}[h]
\centering
\centerline{\includegraphics[height=5cm]{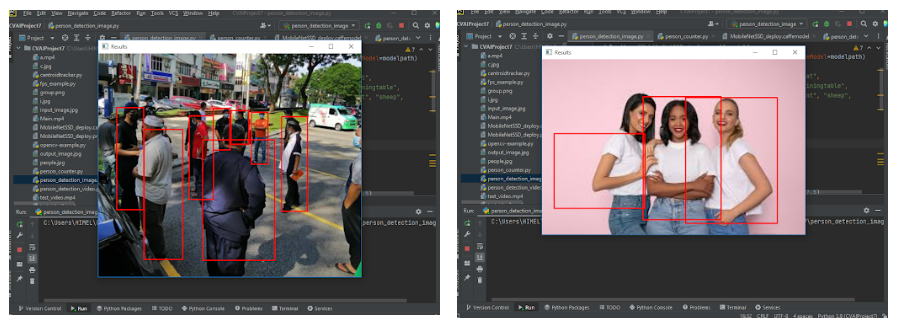}}
\caption{Crowd detected image in street and Crowd detected image of girls group}
\label{fig:r4}
\end{figure*}

\textbf{People detection from Image:} 
Here in Fig.\ref{fig:r4} we can see, our program can detect people easily from images.

\begin{figure*}[h]
\centering
\centerline{\includegraphics[height=5cm]{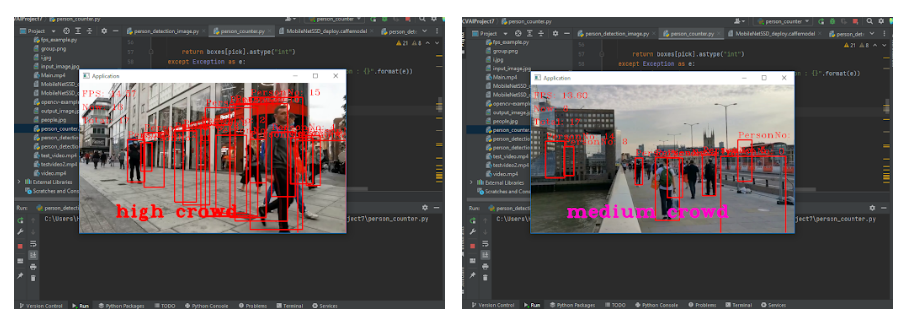}}
\caption{Crowd detected and level estimation(high and medium) video of street}
\label{fig:r5}
\end{figure*}

\textbf{People counter and the whole Monitoring:}
Here in Fig.\ref{fig:r5} we can see our program can count FPS, the current number of people, and the total number of people. Moreover, depending on the current number of people it also shows the estimation level of the crowd as, “High crowd”, “Medium crowd”, or “Normal crowd”.
\newline

\textbf{Sequential Model representation}
\newline
\begin{figure*}[!h]
\centering
\centerline{\includegraphics[height=10cm]{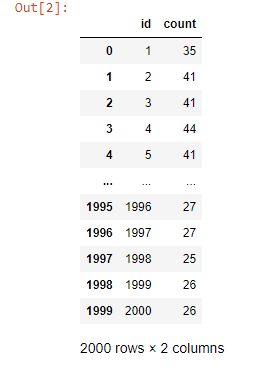}}
\caption{Total dataset elements}
\label{fig:r3}
\end{figure*}
In Fig.\ref{fig:r3} we can see the total number of rows and columns of our dataset.

\begin{figure*}[!h]
\centering
\centerline{\includegraphics[height=6cm]{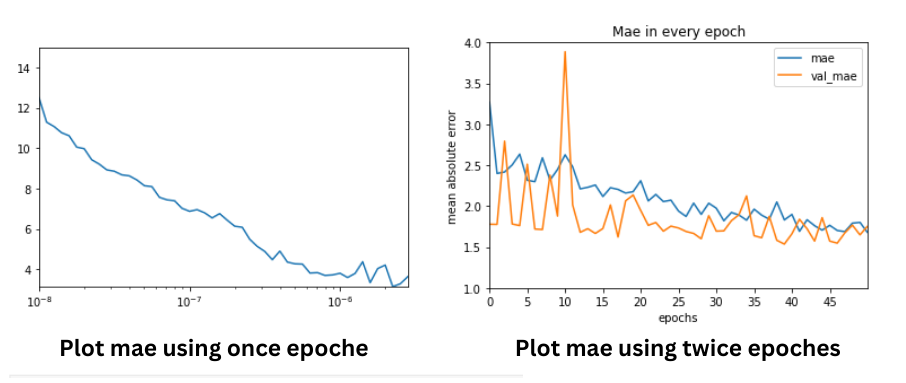}}
\caption{Plot mae using one and two epoches}
\label{fig:r6}
\end{figure*}
Here from Fig.\ref{fig:r6}, we can see plot mae using once epoche and compare the Mean absolute error using twice epochs.

\begin{figure*}[!h]
\centering
\centerline{\includegraphics[height=10cm]{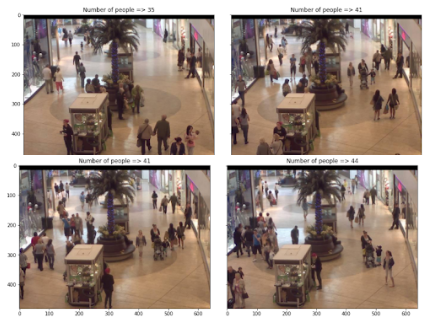}}
\caption{Program is predicting the number of people with figure}
\label{fig:r7}
\end{figure*}
In Fig.\ref{fig:r7} we can see this system is predicting the number of people with the figures.

\begin{figure*}[!h]
\centering
\centerline{\includegraphics[height=5cm]{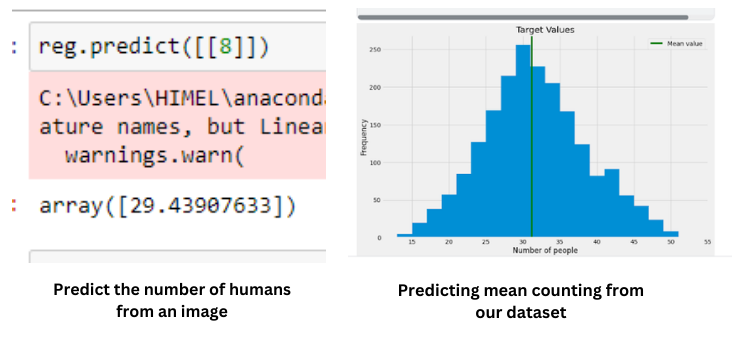}}
\caption{ Predict from one variable and predict the mean counting.}
\label{fig:r8}
\end{figure*}
Here in Fig.\ref{fig:r8} it shows the prediction from index number of figure and predicts the mean counting using bar chart.

\textbf{Linear Regression model}
\newline

\begin{figure*}[h]
\centering
\centerline{\includegraphics[height=2cm]{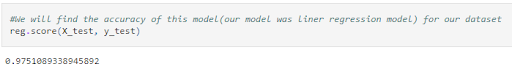}}
\caption{Accuracy rate}
\label{fig:r9}
\end{figure*}

In Fig.\ref{fig:r9} we can see the accuracy rate of our system.

\begin{figure*}[h]
\centering
\centerline{\includegraphics[height=6cm]{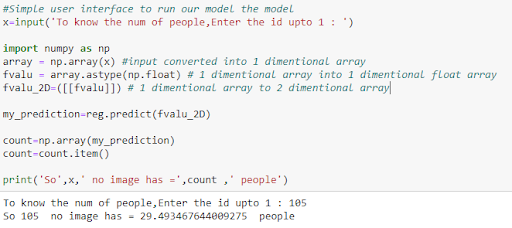}}
\caption{ predict the total number of people from any image}
\label{fig:r10}
\end{figure*}

Here in Fig.\ref{fig:r10}, we can see it can predict the total number of people from any image no.

\subsection{Discussion}
The procedure of crowd density level estimation involves the estimation of the total number of people in a  place. This is done by first analyzing the available information on the area in question such as the size, population density, and type of activity occurring in the area. This can be done through direct observations of individuals, cameras, or other sensing technologies. Once the number of people is counted, the crowd density level is estimated. It provides an estimation level of crowd present in that place. The result of the estimation can then be used to make decisions about the area, such as whether additional resources are needed to manage the crowd or if the area should be limited for safety reasons.
\newline
  
  \textbf{Step 1:}
  We have learned about the basics of ML and Python languages, datasets, open-cv, etc., and also installed the dataset and packages. We have also installed Anaconda and Python.
  \newline
  
  \textbf{Step 2:}
  Then we made a Python file and worked on image detection and wrote something on that figure.
  \newline
  
  \textbf{Step 3:}
  Then we worked on the video. Here we downloaded a video and to detect Human use used the Caffe model. Then import the model and make a Python program that can detect the crowd and give them a unique id for tacking. Here we updated all the crowd counting, for example, Now the total number of people on screen, the total number of people in this video till then, and FPS.
  \newline
  
  \textbf{Step 4:}
  At last, our project also shows the density level according to the counting of the people.
  \newline
  
  \textbf{Step 5:}
  We can do this detection using a webcam also. 
  \newline
  
  \textbf{Step 6:}
  Then we downloaded the dataset “crowd-counting” from Kaggle and did some evaluation using that dataset. Here we also show the graphical representation of the dataset.
  \newline

So here we can see from our project, our project runs well from images, videos, and webcams. It can detect FPS, current counting, and total counting. Also, it can detect density levels so we can easily understand whether a place is overcrowded or not.     
Then we worked with the dataset. We can see the accuracy level and various graphical representations of our project. Also, we showed how to predict the counting. Here is also the mean counting of our dataset. We have used two algorithms to solve it. So we can say, our project is well-shown and well-evaluated.

\section{Conclusion and Future Work}
The research of this study provided proof that crowd density level estimation can be almost accurate, user-friendly, cost-effective, and reliably accomplished using machine learning algorithms and various models. Crowd Density Level Estimation is a user-friendly project to predict crowd detection in various locations, such as occasions, various public concerts, or public transportation. It can help us to be aware of overcrowding by providing better safety and handling the overall crowd management. In this project, we calculate the FPS, total number of people, and current number of people and estimate the crowd level. In today's life, there are various increasing requirements for these types of systems, which can make our life easier, more comfortable, safe, and less need of human attachment. Though our system has many useful and effective features, this system is not free of limitations. In the future, we will train our system with more models. Also, we can research at a more accurate level using larger datasets and models. Additionally, another potential area of research could be to increase the speed for early estimation.

\bibliographystyle{abbrv}
\bibliography{mybib}

\end{document}